\documentclass[aps,twocolumn, superscriptaddress]{revtex4-1}
\newcommand{\beq}{\begin{equation}}
\newcommand{\eeq}{\end{equation}}
\usepackage{graphics}
\usepackage{amsmath}
\usepackage{amsfonts}
\usepackage{amssymb}
\usepackage{graphicx}
\usepackage{color}
\usepackage[normalem]{ulem}

\begin{document}

\title{Exchange dependent ultrafast magnetization dynamics in Fe$_{1-x}$Ni$_{x}$ alloys}

\author{Somnath Jana}
 \affiliation{Department of Physics and Astronomy, Uppsala University, Box 516, 75120 Uppsala, Sweden}
\author{Ronny Knut}
 \affiliation{Department of Physics and Astronomy, Uppsala University, Box 516, 75120 Uppsala, Sweden}
\author{Erna K. Delczeg-Czirjak}
 \affiliation{Department of Physics and Astronomy, Uppsala University, Box 516, 75120 Uppsala, Sweden}
\author{Rameez S. Malik}
 \affiliation{Department of Physics and Astronomy, Uppsala University, Box 516, 75120 Uppsala, Sweden}
\author{Robert Stefanuik}
 \affiliation{Department of Physics and Astronomy, Uppsala University, Box 516, 75120 Uppsala, Sweden}
\author{Raghuveer Chimata}
 \affiliation{Department of Physics and Astronomy, Uppsala University, Box 516, 75120 Uppsala, Sweden}
 \author{Dibya Phuyal}
 \affiliation{Department of Physics and Astronomy, Uppsala University, Box 516, 75120 Uppsala, Sweden}
\author{Venkata Mutta}
 \affiliation{Department of Physics and Astronomy, Uppsala University, Box 516, 75120 Uppsala, Sweden}
\author{Serkan Akansel}
\affiliation{ Department of Engineering Sciences, Uppsala University, Box 534, 75121 Uppsala, Sweden}
\author{Daniel Primetzhofer}
 \affiliation{Department of Physics and Astronomy, Uppsala University, Box 516, 75120 Uppsala, Sweden}
\author{Martina Ahlberg}
\affiliation{Department of Physics, University of Gothenburg, 412 96 Gothenburg, Sweden}
\author{Johan S\"oderstr\"om}
 \affiliation{Department of Physics and Astronomy, Uppsala University, Box 516, 75120 Uppsala, Sweden}
\author{Johan {\AA}kerman}
\affiliation{Department of Physics, University of Gothenburg, 412 96 Gothenburg, Sweden}
\affiliation{Department of Applied Physics, School of Engineering Sciences, KTH Royal Institute of Technology, 164 40 Kista, Sweden}
\author{Peter Svedlindh}
\affiliation{ Department of Engineering Sciences, Uppsala University, Box 534, 75121 Uppsala, Sweden}
\author{Olle Eriksson}
 \affiliation{Department of Physics and Astronomy, Uppsala University, Box 516, 75120 Uppsala, Sweden}
\author{Olof Karis}
 \affiliation{Department of Physics and Astronomy, Uppsala University, Box 516, 75120 Uppsala, Sweden}

%\date{\today}

\begin{abstract}

Element specific ultrafast demagnetization was studied in Fe$_{1-x}$Ni$_{x}$ alloys, covering the concentration range between $0.1<x<0.9$.  For all compositions, we observe a delay in the onset of Ni demagnetization relative to the Fe demagnetization. We find that the delay is correlated to the Curie temperature and hence also  the exchange interaction. The temporal evolution of demagnetization is fitted to a magnon diffusion model based on the presupposition of enhanced ultrafast magnon generation in the Fe sublattice. The spin wave stiffness extracted from this model correspond well to known experimental values.
\end{abstract}

\maketitle

Since the first observation of ultrafast quenching of Ni magnetization after an optical excitation\cite{Beaurepaire1996}, the main objective in the field of ultrafast demagnetization is to determine the channels of angular momentum dissipation from the spin system. To understand the microscopic mechanism involved in ultrafast demagnetization, a diverse set of theoretical models have been employed along with numerous experimental techniques~\cite{Dalla2007,Bigot2009,Schellekens2013,Battiato2012,Zhang2000}. Elliott-Yafet spin-flip processes through electron-phonon scattering can transfer the angular momentum from the spin to the lattice via the spin-orbit interaction, which is often used to describe ultrafast demagetization\cite{Mueller2011,Roth2012,Carpene2008,Michael2009,Koopmans2010}. Observation of simultaneous reduction in both the spin and the orbital  moment, suggest that any angular momentum that is transferred to the orbital part, is relocated to the lattice faster than the experimental temporal resolution ($\sim100$ fs)~\cite{Stamm2007}. 
%Angular momentum transfer between electron spins and photons via the spin-orbit interaction\cite{Zhang2000,Bigot2009} fails to explain the sizeable amount of demagnetization found after the duration of the pump pulse. 
Diffusion of spin-polarized electrons to the substrate has been proposed as a mechanism for the dissipation of angular momentum from the probed region\cite{Battiato2012}. Studies of Fe/spacer/Ni tri-layers indicate that both spin-flip scattering and spin diffusion are present, where the respective contribution from these processes depend on the conducting property of the nonmagnetic spacer layer\cite{Turgut2013}. Calculations have been performed to estimate the demagnetization rate on the basis of spin-phonon scattering~\cite{Essert2011,Carva2011,Carva2013,Illg2013,Mueller2013}, but none  can successfully reproduce the large demagnetization observed experimentally. Carpene et al.~\cite{Carpene2008} argued that the demagnetization is determined by magnon emission, where the electron-magnon scattering can transfer angular momentum from the spin to the orbital magnetic moment, which is immediately quenched by the crystal field. However, it was found that the calculated demagnetization rate for Fe and Ni, when considering only magnon emission, is too small compared to the experimental values~\cite{Illg2013}. Furthermore, Haag et al.~\cite{Haag2014} suggested that neither electron scattering from phonons nor electron scattering from magnons can independently explain experimental demagnetization rates, while it could be possible when combining both processes. Various probing techniques show indirect evidence of ultrafast magnon emission~\cite{Schmidt2010,Atxitia2010}, however, direct evidence is lacking.

Table top ultra-short extreme ultraviolet (XUV) sources provides element selective magnetization measurements by probing the M-edges of magnetic transition metals. The elemental specificity can be used for separating the dynamical response between constituents of an alloy or in layered magnetic structures. In recent work, Ref.~\cite{Mathias2012}, element resolved demagnetization in Permalloy (Py) revealed a delay in the onset of Ni demagnetization relative to that of Fe, which has been confirmed by others~\cite{Jana2017,PhysRevB.90.180407}. 
Furthermore, when Py was alloyed with 40\% Cu it was found that the observed delay increased by a factor inversely proportional to the exchange interaction. It was suggested that only the Fe sublattice participate in the direct demagnetization process while Ni demagnetizes through the exchange interaction with Fe. In this work, we have measured the element specific demagnetization in a series of Fe$_{1-x}$Ni$_{x}$ alloys with six different Ni concentrations. A concentration dependent variation in the relative delay is observed, which is found to be inversely proportional to the Curie temperature. The experimental demagnetization is fitted using a model describing preferential magnon scattering in the Fe sublattice\cite{Ronny2017} which provides values for the spin wave stiffness ($D_{spin}$) in each alloy composition. These extracted values correspond well to calculated ab-initio values, values derived from the Bloch T$^{3/2}$ law, and literature values of the exchange stiffness obtained from neutron scattering \cite{Vaz2008}.

%\section{Experimental and theoretical methods}
A sample series with the geometry, Si-substrate/Ta(2 nm)/Fe$_{1-x}$Ni$_{x}$(20 nm)/Ta(2 nm) and nominal concentrations of $x =$ 0.0, 0.2, 0.3, 0.4, 0.6, 0.7, 0.8, 1.0, were produced by magnetron sputtering at a base pressure of 10$^{-8}$ Torr, Ar pressure of 3 mTorr and flow 30 sccm. Rutherford back scattering was performed to determine the exact compositions and thicknesses of the alloys, giving 20\%, 30\%, 36\%, 58\%, 68\%, 78\% Ni and 18.4 nm, 17.7 nm, 18.2 nm, 17.4 nm, 16.6 nm, 16.6 nm, respectively.
All the samples were characterized by X-ray diffraction (XRD), which provides structural information (see SI). XRD confirms a single phase for all samples, with bcc structures for x = 0.0, 0.2, 0.3, 0.4 and fcc structures for x = 0.6, 0.7, 0.8, 1. Static temperature dependent magnetic measurement was performed using a SQUID magnetometer in a field of 1.0 T to estimate the spin wave stiffness.
%%%%%----------------------THEORY-----------EMTO------START---------%%%%%%%%%%%%%%%%%%%%%
Concentration dependent spin wave stiffness was obtained by means of first principles calculations. The magnetic exchange integrals were calculated by means of density functional theory (DFT) \cite{Hohenberg1964,Kohn1965} using the LKAG-formalism \cite{Liechtenstein1984} as implemented in the Exact Muffin-Tin Orbitals-Full Charge Density (EMTO-FCD) method \cite{Vitos2007} using the local density approximation \cite{Perdew1992}. Chemical disorder was treated by means of the coherent potential approximation \cite{Soven1967, Gyorffy1972}. The Green's function was calculated for 16 complex energy points distributed exponentially in a semicircular contour. The $s$, $p$, $d$ and $f$ orbitals ($l_{max}$=3) were included in the EMTO basis set. The one-center expansion of the full charge density was truncated at $l^h_{max}$=8. To obtain the accuracy needed for the exchange integral calculation we used 3000 $k$-points in the irreducible wedge of the Brillouin zone. The exchange integrals were calculated up to 3.5 times the lattice parameter distance. All the calculations was carried out using the experimental lattice parameters ($a$). The spin wave stiffness is calculated using the Uppsala atomistic spin dynamics code (UppASD) \cite{UppASD} in terms of total exchange constant $J_{0j}$ as \cite{Pajda2001},
 \begin{equation}\label{D2}
\begin{split}
D_{spin} & =lim_{\eta \to 0} D(\eta) \\
 & =\frac{2\mu_{B}}{3M}\sum_{0<R_{0j}<R_{max}}J_{0j}R^{2}_{0j}exp(-\eta R_{0j}/a)
\end{split}
\end{equation}
where $R_{0j}=|{\bf R}_{0j}|$ denotes the distance between atomic sites. $R_{max}$ must be considered carefully in the calculations since the exchange interactions have long-ranged oscillatory RKKY-like behaviour. 
Also, the quantity $\eta$ plays an important role to make the sum over $R_{0j}$ absolutely convergent.
%%%%%%%%%%%%%%%%%%%----------------------THEORY-----------EMTO------END---------%%%%%%%%%%

\begin{figure}[tbp]
\includegraphics[width=0.99\columnwidth]{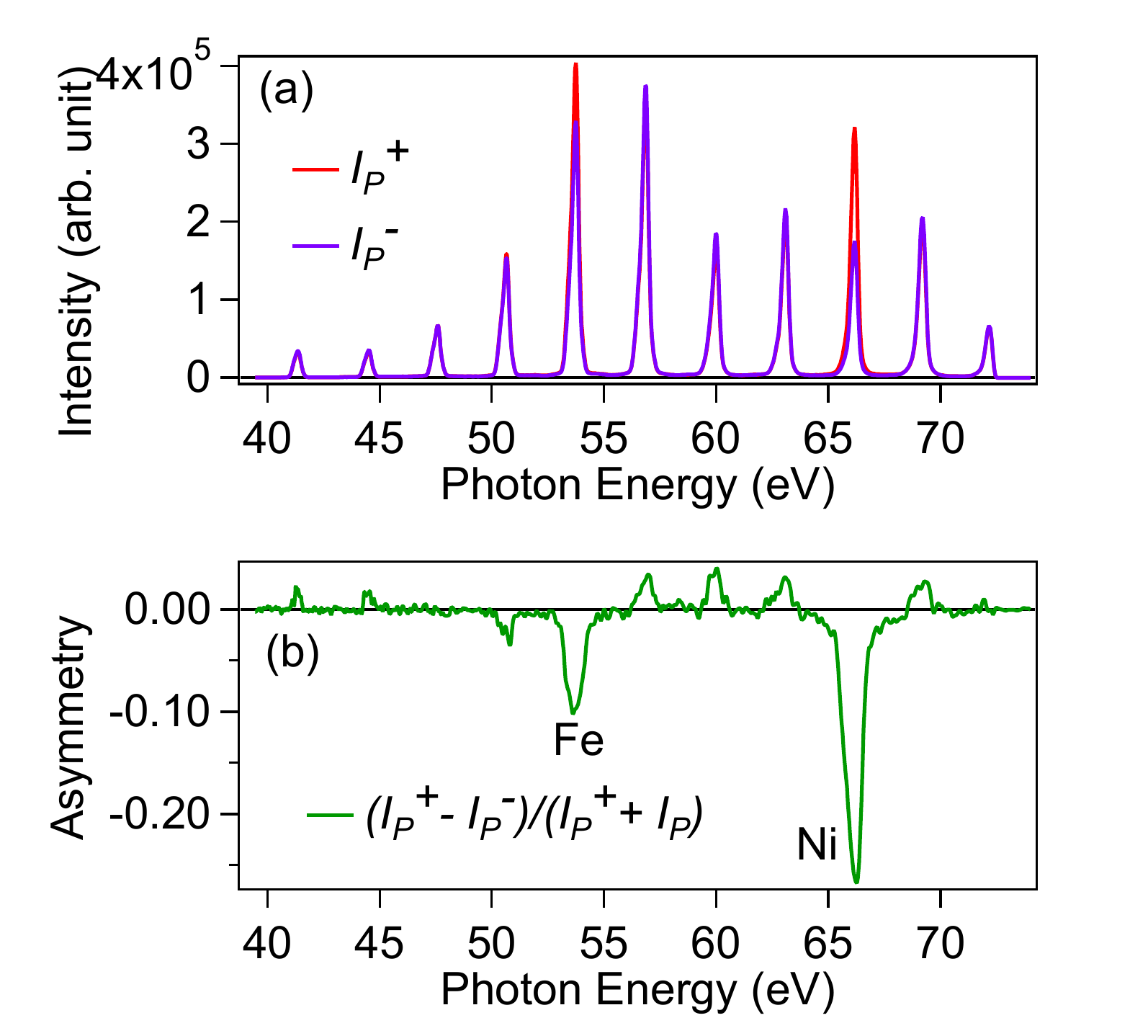}
\caption{(Color Online) (a) Reflected XUV spectra collected for two opposite in plane magnetization directions for a non-demagnetized state of Fe$_{0.22}$Ni$_{0.78}$, (b) asymmetry spectra obtained from (a).\label{Fig_Spectrum}}
\end{figure}

The experimental setup for studying ultrafast demagnetization consists of a femtosecond infrared (IR) laser (800 nm, 35 fs, 0.8 mJ, 10 KHz), where 80\% of the light is focused into a He filled gas cell for high harmonic generation (HHG). This produces ultrashort XUV pulses that retain the polarization and coherence of the driving IR \cite{Corkum1993}. The p-polarized XUV pulses, with energies ranging from 35 to 72 eV (see Fig.\ \ref{Fig_Spectrum}) is focused on the sample at an incidence angle of 45 degrees. A grating spectrometer spectrally resolves the XUV on a position sensitive multichannel plate detector. The magnetization is probed by calculating the asymmetry parameter ($A(E)$), defined by the normalized difference spectra of the reflected intensities ($I_p^{\pm}(E)$) measured for opposite magnetic field directions, applied parallel to the sample surface and perpendicular to the plane of the incidence. 
\begin{equation}\label{eq:asymdiff}
A(E) = \frac{I_p^+(E) - I_p^- (E)}{I_p^+(E) + I_p^- (E)} \:
\end{equation}

A fraction of the remaining 800 nm light of the laser is used as a pump-pulse for demagnetizing the sample. Since the pump and probe pulses originate from the same initial laser pulse, the temporal jitter is practically eliminated.
An optical delay stage is employed in the pump line to introduce a controllable delay between the IR pump and the XUV probe. More details of the sample geometry, spectrometer and the measurement protocol are described in Ref.~\cite{Jana2017}. The HHG setup is described in detail in Ref. \cite{Plogmaker2015}.

%\section{Results and discussions}
A typical spectrum is shown in Fig.\ \ref{Fig_Spectrum}(a), obtained with two opposite magnetic field directions for Py (Fe$_{0.22}$Ni$_{0.78}$). The odd harmonics are well separated from each other by 3.1 eV, so the asymmetry can be measured as a function of harmonic energy without ambiguity. 
The corresponding asymmetry spectra is shown in Fig.\ \ref{Fig_Spectrum}(b). The asymmetry is strongly enhanced near 54 eV and 62 eV, corresponding to the Fe and Ni M$_{2,3}$ absorption edges, respectively. Since, to a first order approximation the asymmetry is proportional to the magnetization\cite{PhysRevX.2.011005}, the elemental magnetization can be obtained from the asymmetry at the Fe and Ni M$_{2,3}$ absorption edges. The simultaneous recording of the whole spectrum enables detecting relative dynamics of the constituent elements without experimental artifacts from jitter and drift.

\begin{figure}[tbp]
\includegraphics[width=0.99\columnwidth]{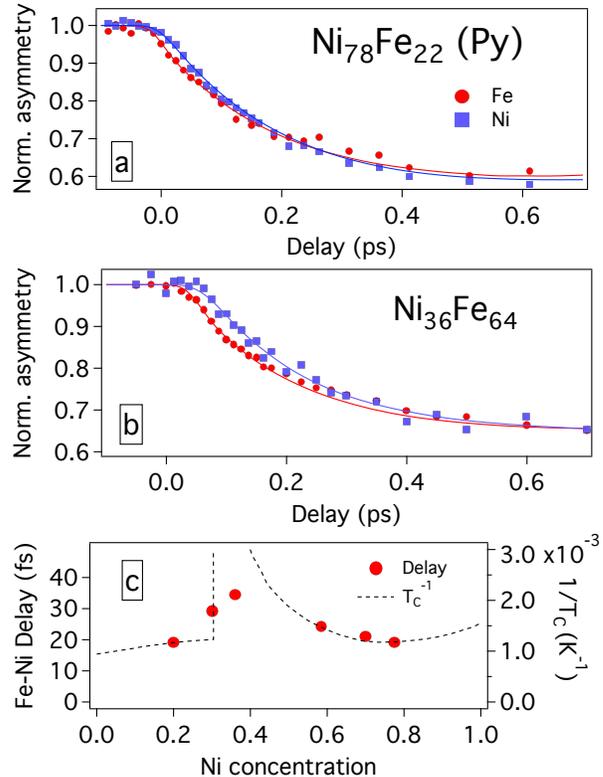}
\caption{(Colour Online) Asymmetry obtained at Fe $M_{2/3}$ ($\sim$54 eV) and Ni $M_{2/3}$ ($\sim$66 eV) edges are plotted (solid circles for Fe and solid squares for Ni) with the pump-probe delay in panel (a) and (b) for compositions x = 0.78 and 0.36 respectively. The lines through the experimental data points are the fitted curves using the magnon diffusion model. (c) The inverse transition temperature \cite{Kakehashi1990} and the obtained delay in Ni demagnetization relative to that of Fe are plotted versus the compositional variation. \label{Fig_DemagCurve}}
\end{figure}

Pump-probe measurements were performed for all samples. In Fig.\ \ref{Fig_DemagCurve}(a) and (b), the normalized asymmetry at the Fe and Ni M$_{2,3}$ edges are plotted as a function of the delay between the pump and probe for two compositions (other compositions are presented in the SI). All compositions display a delay in the onset of Ni demagnetization relative to Fe. The delay is practically independent of the pump fluence (see SI), however, the pump fluence was maintained such that the maximum demagnetization is about 30-40\% for all samples. The delay is quantified by shifting the Fe demagnetization and calculating the root mean square (RMS) of the difference between Fe and Ni demagnetization within a 150 fs window around the onset of demagnetization. The shift that provides the lowest RMS is used as the delay between Fe and Ni demagnetization. The results are shown in Fig.\ \ref{Fig_DemagCurve}(c) for the different compositions. For a Fe concentration of 22\% the delay is 19 fs, which is close to what has been reported in earlier studies ~\cite{Mathias2012, Jana2017} for Py. Starting from a high Ni concentration the delay first shows an increase, peaks around 40\% Ni, and then decreases for lower Ni concentrations. The inverse of the calculated Curie temperature~\cite{Kakehashi1990}  is plotted along the compositions on the right axis of Fig.\ \ref{Fig_DemagCurve}(c), indicating a correlation between the average exchange interaction of the samples and the demagnetization delay between Fe and Ni. A reference sample consisting of elemental Fe and Ni grown on the same substrate exhibits no delayed demagnetization response, indicating that the observed effect is a consequence of alloying (see SI).

It was suggested by Knut et al.\cite{Ronny2017} that the delayed response in the Ni sublattice is due to ultrafast magnon generation that occurs dominantly in the Fe sublattice. In this case, the spatially inhomogeneous magnon distribution, at short timescales, would diffuse and increase the magnon population in the Ni sublattice at longer timescales. This would be manifested as a delayed demagnetization in the Ni sublattice. Note that the demagnetization is still driven by some other process, such as EY-spin flips, since the magnon generation is considered to be primarily spin-conserving. In Fig.\ \ref{Fig_DemagCurve}(a) and (b) the solid lines represents fits using the model by Knut et al.\cite{Ronny2017}. In this model, a 1D lattice with 1024 randomly distributed Fe and Ni atoms is generated. The inhomogeneous magnon distribution is generated by a non-uniform exchange scattering in the alloy, where the ratio of $sd$-exchange interaction ($J^{sd}_{Fe,Ni}$) between Fe and Ni is fixed at $J^{sd}_{Fe}/J^{sd}_{Ni}=2.2$.
 The magnon probability distribution is then described by a spatially dependent amplitude of the magnon wave function. %, where the phase of the wave function is determined by the magnon wave number. 
The magnon wave number distribution is described by the Bose-Einstein distribution  at equilibrium. However, at short timescales after a laser excitation it is strongly weighted towards short lifetime magnons with high wave numbers.  Here we have used a wave number distribution that corresponds to 100 fs after the laser excitation, since it represents the time window where Ni exhibit a delay relative to Fe.
Other parameters are the same as given in Knut et al.\cite{Ronny2017}. The magnon diffusion and hence the delay of Ni demagnetization depends on the spin wave stiffness.
\begin{figure}[tbp]
\includegraphics[width=0.99\columnwidth]{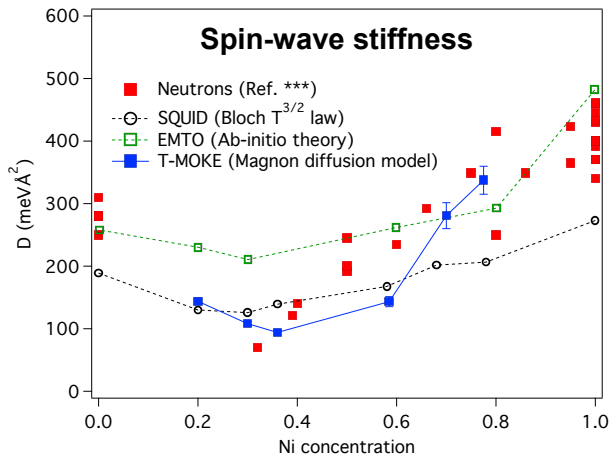}
\caption{(Colour Online) Spin wave stiffness obtained from the element specific demagnetization curves, saturation magnetization, theoretical calculation and neutron scattering results~\cite{Vaz2008} are plotted with the Ni concentrations in Fe$_{1-x}$Ni$_{x}$ alloy. \label{Fig_SpinnWaveStiff}}
\end{figure}
In Fig.\ \ref{Fig_SpinnWaveStiff}, the fitted value of the spin wave stiffness (squares) is plotted versus the Ni concentration together with reference data measured by neutron scattering experiment (circles).
Furthermore, the spin wave stiffness was also extracted by measuring the temperature dependent saturation magnetization $m_s(T)$ of the samples which is fitted to the Bloch $T{^{3/2}}$ law, \([m_s(0)-m_s(T)]\propto ({k_{B}T/D_{spin}})^{3/2}\) in the temperature range of 10-150 K, where T is temperature, and $k_{B}$ is Boltzman constant. For details see SI and also Ref. \cite{Nembach2015}. Finally, we also show ab-initio calculated values of the spin-wave stiffness (triangles). 
The derived values of $D_{spin}$ from the simulation of inhomogeneous magnon generation are similar to the values obtained by other methods. Note that the fitted values are sensitive to the ratio of exchange scattering between Fe and Ni, which was chosen to provide a reasonable correspondence to neutron scattering data for Fe$_{0.22}$Ni$_{0.78}$.
%However, the  fitted value of $D_{spin}$ is sensitive to the asymmetry of magnon generation between Fe and Ni atoms, where we have used the ratio of 4:1 for all samples.
 The calculated magnon scattering rate in Fe by Haag et al.\cite{Haag2014} is an order of magnitude higher than in Ni, suggesting that the ratio of 2.2 in Fe-Ni alloys is a reasonable value. More interesting is the general trend of $D_{spin}$ which is similar for all the different methods of extracting the spin wave stiffness. It is clear that the spin wave stiffness decreases when pure Ni is alloyed with Fe. The minimum value is observed around the Invar transition (35\% Ni), after which it increases for higher Fe concentrations.

%\section{Conclusion}
In conclusion, element specific demagnetization dynamics in Fe$_{1-x}$Ni$_{x}$ alloys were measured using the time-resolved T-MOKE technique with ultrashort XUV pulses. A delay in the onset of the Ni demagnetization, relative to that of Fe, was observed for all studied compositions. This delay is inherent to Fe-Ni alloys, and is not observed for elemental Fe and Ni. The delay was found to be correlated with the inverse of the Curie temperature across the whole compositions range. We find that the Fe and Ni demagnetizations are well described by a ultrafast inhomogeneous magnon scattering. By adopting a higher magnon generation rate at Fe-sites compared to Ni-sites, we extract the spin wave stiffness for all compositions which are in excellent agreement with the values obtained from neutron scattering measurement, static magnetization data and ab-initio calculations. 

\section{Acknowledgement}
We gratefully acknowledge Hans Nembach and Tom Silva for useful discussions. E. K. Delczeg-Cz. acknowledges the Swedish National Infrastructure for Computing (SNIC) for computational resources and STandUP, eSSENCE, SSF, VR for financial support.

%Also include:
%Carl Tryggers Foundation 
%Knut and Alice Wallenbergs Foundatio
%\bibliographystyle{unsrt}
% \bibliography{Fe1-xNix_2}


\end{document}